\documentclass[sigconf]{acmart}

\AtBeginDocument{%
  \providecommand\BibTeX{{%
    \normalfont B\kern-0.5em{\scshape i\kern-0.25em b}\kern-0.8em\TeX}}}

\setcopyright{acmcopyright}
\copyrightyear{2020}
\acmYear{2020}
\setcopyright{acmcopyright}\acmConference[ICAIF '20]{ACM International Conference on AI in Finance}{October 15--16, 2020}{New York, NY, USA}
\acmBooktitle{ACM International Conference on AI in Finance (ICAIF '20), October 15--16, 2020, New York, NY, USA}
\acmPrice{15.00}
\acmDOI{10.1145/3383455.3422523}
\acmISBN{978-1-4503-7584-9/20/10}

\begin{document}

\title{Analysis of the impact of maker-taker fees on the stock market using agent-based simulation}
\author{Isao Yagi}
\orcid{0000-0003-0119-1366}
\authornotemark[1]
\affiliation{%
  \institution{Kanagawa Institute of Technology}
  \streetaddress{Shimoogino 1030}
  \city{Atsugi}
  \state{Kanagawa}
  \country{Japan}
}
\email{iyagi2005@gmail.com}

\author{Mahiro Hoshino}
\affiliation{%
  \institution{Kanagawa Institute of Technology}
  \city{Atsugi}
  \state{Kanagawa}
  \country{Japan}
}

\author{Takanobu Mizuta}
\affiliation{%
  \institution{SPARX Asset Management Co., Ltd.}
  \city{Minato}
  \state{Tokyo}
  \country{Japan}
}
\email{mizutata@gmail.com}

\renewcommand{\shortauthors}{Yagi et al.}

\begin{abstract}
Recently, most stock exchanges in the U.S.\ employ maker-taker fees, in which an exchange pays rebates to traders placing orders in the order book and charges fees to traders taking orders from the order book. Maker-taker fees encourage traders to place many orders that provide market liquidity to the exchange. However, it is not clear how maker-taker fees affect the total cost of a taking order, including all the charged fees  and the market impact. In this study, we investigated the effect of maker-taker fees on the total cost of a taking order with our artificial market model, which is an agent-based model for financial markets. We found that maker-taker fees encourage market efficiency but increase the total costs of taking orders.
\end{abstract}




\maketitle

\section{Introduction}
These days, most equities exchanges in the U.S.\ are discussing maker-taker fees, in which an exchange pays rebates to makers who are traders placing orders in the order book and charges fees to takers who are traders taking orders from the order book \cite{Do12,BCJ16}. 
The merits of maker-taker fees are that market liquidity is provided for taking orders, improving liquidity may increase the exchange's trading share, the bid-ask spread can be decreased by maker orders, and the market can be expected to be more efficient. 
There have been some previous reports on maker-taker fees \cite{FKK13,CVV19,Ro19};
however, it remains unclear how maker-taker fees affect the total cost of a taking order, including all the charged fees and the market impact. 

In this research, we investigated the effect of maker-taker fees on the market with our artificial market model, which is an agent-based simulation model for financial markets \cite{CIP09,CCD12,MIYY14}. 

One method for dealing with situations that defy analysis by previous empirical research methods is an artificial market. 
In this case, it is difficult to investigate in empirical research the impact on the market caused by the amount of maker-taker fees that are not actually used, but if artificial markets are used, the effect of these fees on the market can be easily understood.
Each agent is assigned a specific trading (i.e., buying or selling) rule and is then set to trade financial assets as an investor. The market can be observed in order to determine how the agents behave. Then, by modeling and incorporating certain restrictions on the market side (e.g., limitations to ensure market stability and efficiency such as a short selling regulation), it is possible to examine how investors behave, as well as what kinds of effect the restrictions have on the market.

We constructed an artificial market with the maker-taker fee structure and analyzed the effect of maker-taker fees, i.e., the total cost of a taker's trade, including the charged fee, on the market from the viewpoints of market impact and volatility.
Furthermore, we checked the relationship between the total cost of a taker's trade and 
market efficiency.

\section{Artificial market model}\label{Artificial market model}
In this research, we constructed an artificial market model on the basis of the artificial market model of Chiarella et al.\ \cite{CIP09} and Yagi et al.\ \cite{YMM19}, because 
they built simple agents and a pricing mechanism which could reproduce the statistical characteristics of the kinds of long-term price fluctuations observed in empirical analyses.
Moreover, algorithm agents \cite{MKKMI15} and a position-based market maker \cite{KMHI14} were added to our model.
The normal agents used in the basis models \cite{CIP09, YMM19} correspond to both takers and makers in real financial markets, where the algorithm agents correspond to takers and the position-based market makers correspond to makers.

Only one risk asset is available for trading. Hereinafter, we refer to risk assets as simply assets, whereas we refer to non-risk assets as cash, because non-risk assets are equivalent to cash in the proposed artificial market. 

The numbers of normal agents and algorithm agents are $n$ and $m$ (where $1 \leq m \leq n$), respectively. Each of the normal agents $j=1, \ldots, n$ places an order in sequence. After the final agent, agent $n$, has placed an order, the first agent, agent $1$, places the next order. Each time $\lfloor n/m \rfloor$ normal agents place orders, the algorithm agent places one order. Each of the algorithm agents $k=1, \ldots, m$ also places an order in sequence. There is one position-based market maker, which places a sell order and a buy order before either the normal agents or algorithm agents place orders.

The time $t$ increases by 1 each time a normal agent or an algorithm agent places an order.
Thus, the process moves forward one step even when a trade does not occur and this new order is placed in the order book. However, time $t$ does not advance in the position-based market maker order.

The mechanism for determining the price in this model being a continuous double auction (continuous trading session) means that if there are sell (buy) order prices in the order book that are lower (higher) than the buy (sell) order price of the agent, then the agent's order is immediately matched to the lowest sell order (highest buy order) in the order book. We refer to this as a market order.
If there are no such orders in the order book, then the order does not match any other order and remains in the order book. We refer to this order as a limit order.

The remaining orders in the order book are canceled at time $t_c$ (order effective period) after the order was placed. 

The tick size, which is the minimum unit for price, is $\Delta{P}$, and when orders are sell orders, fractional values smaller than $\Delta{P}$ are rounded up. On the other hand, when orders are buy orders, they are rounded down.
Each agent sends one order each time. An agent can possess an unlimited amount of assets as the quantity of cash held by the agent is set to infinity. Agents can also short sell.

\subsection{Maker-taker fee structure}\label{Maker-taker fee structure}
One of the exchange's revenues is charging traders for the services that process their trades. Maker-taker fees provide a benefit for exchanges in that the exchanges can profit from the difference between the taker fee and the market maker rebate as follows: 
\begin{equation}
	R_{EX}=C_T-R_M.
	\label{eq1}
\end{equation}
Let $R_{EX}$, $C_T$, and $R_M$ be the fee that the exchange receives, the taker fee, and the market maker rebate, respectively. In this study, $R_{EX}$ is fixed ($R_{EX} = 0.100 \%$), since $R_{EX}$ has essentially no effect on the relationship between the total cost of a taker's trade, including the taker fee, and market efficiency.
When $R_M$ is positive, it means that a market maker receives the market maker rebate. When $R_M$ is negative, it means that a market maker pays the market maker rebate. 
When $C_T$ is positive, it means that a taker pays the taker fee.
When $C_T$ is negative, it means that a taker receives the taker fee. The values of $R_ {EX}$, $R_M$, and $C_T$ are shown as ratios to the fundamental price $P_f$ described later.

\subsection{Normal agent}\label{Normal agent}
Normal agents, who are assumed to be general investors in the real world, are designed for as simple a model which can reproduce the desired stylized facts.
 Normal agents have three trading strategies: fundamental strategy, technical strategy, and noise trading. The first two strategies are implemented in our model because many empirical studies have found that the fundamental strategy, technical strategy, or both were used generally for any market and any time (e.g., Menkhoff et al.\ \cite{MT07}). We also implement noise trading to model objectively investors' desire for a better strategy through trial and error. 

\subsubsection{Order process}\label{Order process}
The order prices of normal agent $j$ by transaction are determined as shown below.
The rate of change of the price expected by agent $j$ at time $t$ (the expected return) ${r_e}_j^t$ is given by

\begin{equation}
	{r_e}^t_j=\frac{1}{{w_1}^t_j+{w_2}^t_j+u_j}\left({w_1}^t_j{r_1}^t_j+{w_2}^t_j{r_2}^t_j+u_j\epsilon_j^t\right),
		\label{exp_return}
\end{equation}
where ${w_i}^t_j$ is the weight of the $i$-th term for agent $j$ at time $t$ and is set according to the uniform distribution between $0$ and $w_{i,max}$ at the start of the simulation and then varied using the learning process described later herein.
Furthermore, $u_j$, the weight of the third term in parentheses, is set according to the uniform distribution between 0 and $u_{max}$ at the start of the simulation and is kept constant thereafter.

The first term in parentheses on the right-hand side of Eq.~(\ref{exp_return}) represents the fundamental strategy. The form of the term indicates that an agent expects a positive (negative) return when the market price is lower (higher) than the fundamental price.
The term ${r_1}^t_j$ is the expected return of the fundamental strategy for agent $j$ at time $t$ and is given by ${r_1}^t_j = \log(P_f / P^{t-n})$, where 
$P_f$ is the fundamental price, which is constant over time, and $P^t$ is the market price at time $t$. The market price is set to the most recent price at the time if no trading is occurring. The initial market price is set to the fundamental price, i.e., $P^0=P_f$. 

The second term represents the technical strategy. The form of this term indicates that an agent expects a positive (negative) return when the historical return is positive (negative). 
Here, ${r_2}^t_j$ is the expected return of the technical strategy for agent $j$ at time $t$ and is given by ${r_2}^t_j = \log (P^{t-n} / P^{t-n-\tau_j})$, where $\tau_j$ is set according to the uniform distribution between $1$ and $\tau_{max}$ at the start of the simulation.

The third term represents the noise strategy.
Here, $\epsilon_j^t$ is a normally distributed random error with mean zero and standard deviation $\sigma_{\epsilon}$.

Based on the expected return ${r_e}_j^t$, the expected price ${P_e}_j^t$ is found using the following equation: 
\begin{equation}
	{P_e}_j^t = P^{t-1}\exp({r_e}_j^t).
	\label{Pejt}
\end{equation}

The order price ${P_o}_j^t$ is a normally distributed random number with mean ${P_e}_j^t$ and standard deviation $P_{\sigma}^t$, given by 
\begin{equation}
	P_{\sigma}^t = {P_e}_j^t \cdot Est,
\end{equation}
where $Est(0<Est\le1)$ is the variation coefficient of the order price. 
The choice between buying and selling is determined by the relative sizes of the expected price ${P_e}_j^t$ and the order price ${P_o}_j^t$. An agent places a buy order for one share if ${P_e}_j^t~>~{P_o}_j^t$, while an agent places a sell order for one share if ${P_e}_j^t~<~{P_o}_j^t$.

\subsection{Learning process}\label{Learning process}
Previous studies using an artificial market have implemented various kinds of learning processes. For example, agents switch strategies and/or tune their strategy parameters based on their performance, market price, etc.\ \cite{AHLPT96,LM99,NT13}.
The learning process in the present study is implemented to switch the strategy between the fundamental and technical strategies.

We modeled the learning process as follows based on Yagi et al.\ \cite{YMM19}.
For ${r_i}^t_j$, learning is performed by each agent immediately before the agent places an order. That is, when ${r_i}^t_j$ and ${r_l} ^t = \log(P^t/P^{t-t_l})$ are of the same sign, ${w_i}^t_j$ is updated as follows:

\begin{equation}
		{w_i}^t_j{\leftarrow} {w_i}^t_j + k_l|{r_l}^t|q_j^t(w_{i,max} - {w_i}^t_j),
\end{equation}
where $k_l$ is a constant, and $q^t_j$ is set according to the uniform distribution between 0 and 1.
When ${r_i}^t_j$ and ${r_l}^t$ have opposite signs, ${w_i}^t_j$ is updated as follows:
\begin{equation}
		{w_i}^t_j{\leftarrow} {w_i}^t_j - k_l|{r_l}^t|q_j^t{w_i}^t_j.
\end{equation}
Separately from the process for learning based on past performance, ${w_i}^t_j$ is reset with a small probability $m$, according to the uniform distribution between $0$ and $w_{i,max}$.

\subsection{Algorithm agent}\label{Algorithm agent}
Algorithm agents are assumed to be institutional investors who use an algorithmic trading strategy. Algorithmic trading is a process in which a big order is divided into small orders and automatically executed little by little. All algorithm agents always place buy market orders for one share. The purpose of incorporating algorithm agents into our artificial market is to use them to measure the market impact ({\it MI}) described later. 

When there is an order with the best ask price in the order book, an algorithm agent places a buy order at the best ask price plus tick size $\Delta P$. Otherwise, an algorithm agent does not place an order.

\subsection{Position-based market maker}\label{Position-based market maker}
A position-based market maker is assumed to be an institutional investor who takes a market maker strategy, which means the strategy of placing both a buy order and a sell order whose price exceeds the buy order price, to make a profit. Hereinafter, a positon-based market maker is simply called a market maker.

If the previous sell, buy, or both orders of the market maker remain in the order book, the market maker cancels them and places new buy and sell limit orders. 
Generally, a market maker decides their own order price based on the best-bid, the best-ask, and the spread, which is equal to the amount of its own expected return per transaction. However, the order price of the market maker also depends on their position, which means the amount of an asset held by the market maker, as they act to keep their position neutral \cite{NS04,KMHI14}. That is, when the market maker has a long position, which means the agent buys and holds some amount of an asset, their buy and sell order prices are set lower so that their sell order matches an order from normal agents and algorithm agents easier than its buy order. On the other hand, when the market maker has a short position, which means the agent short-sells the asset, their buy and sell order prices are set higher so that their buy order matches an order from normal agents easier than their sell order.

Let the base spread of the market maker and the coefficient of their position (its initial value is set based on Kusaka et al.\ \cite{KMHI14} as $5.0\times10^{-8}$) be $\theta_M$ and $w_M$. Let the best-bid, the best-ask, and the basic order price, buy order price, sell order price at time $t$, and position between time $t$ and $t+1$ of the market maker be ${P_{bb}}^t$, ${P_{ba}}^t$, ${P_{bv,M}}^t$, ${P_{bo,M}}^t$, ${P_{so,M}}^t$, and ${s_M}^t$, respectively. Then, ${P_{bo,M}}^t$, ${P_{so,M}}^t$, and ${P_{bv,M}}^t$ are as follows: 
\begin{align}
	{P_{bo,M}}^t  & = {P_{bv,M}}^t - \frac{1}{2}P_f\cdot \theta_M, \label{PHt} \\
	{P_{so,M}}^t  & = {P_{bv,M}}^t + \frac{1}{2}P_f\cdot \theta_M,	\\
	{P_{bv,M}}^t  & = (1-w_M(s_M^t)^3)\cdot \frac{1}{2}({P_{bb}}^t + {P_{ba}}^t) .	
\end{align}

When the sell (buy) order price of the market maker is lower (higher) than the best-bid (best-ask), the market maker's order becomes a market order. Therefore, if the following conditions are satisfied, the buy and sell order prices of the market maker are changed \cite{KMHI15}. 
That is, if ${P_{bo,M}}^t \geq {P_{ba}}^t$, then
\begin{align}
  {P_{bo,M}}^t & = {P_{ba}}^t - \Delta P, \label{PHt1} \\
  {P_{so,M}}^t & = ({P_{ba}}^t - \Delta P) + P_f\cdot\theta_M.
\end{align}
If ${P_{so,M}}^t \leq {P_{bb}}^t$, then
\begin{align}
  {P_{bo,M}}^t & = ({P_{bb}}^t + \Delta P) - P_f\cdot\theta_M, \label{PHt2} \\
  {P_{so,M}}^t & = {P_{bb}}^t + \Delta P. 
\end{align}

 is set in consideration of the market maker rebate. That is,
\begin{equation}
	\theta_M=Re_M-2R_M,
	\label{basespread}
\end{equation}
where $Re_M$ means the expected return of the market maker per transaction. Note that $Re_M$ is shown as a ratio to the fundamental price $P_f$ and is set to $0.300\%$. The reason why $R_M$ has a factor of 2 in the equation is that the market maker may receive rebates when both a buy limit order and a sell limit order of the market maker are executed.

Table \ref{table1} shows the relationship between $\theta_M$ and $C_T$ by listing their values for various market maker rebates $R_M$ under the above conditions.

\begin{table}[t]
	\caption{Relationship between base spread $\theta_M$ and taker fee $C_T$ with respect to market maker rebate $R_M$ when $Re_M=0.300\%$ and $R_{EX}=0.100\%$}
	\begin{tabular}{c|c|c} 
		$R_M$ & $\theta_M$ & $C_T$\\ \hline\hline
		 -0.100\% & 0.500\% & 0.000\% \\
		 -0.075\% & 0.450\% & 0.025\% \\
		 -0.050\% & 0.400\% & 0.050\% \\
		 -0.025\% & 0.350\% & 0.075\% \\
		 -0.000\% & 0.300\% & 0.100\% \\
		0.025\% & 0.250\% & 0.125\% \\
		0.050\% & 0.200\% & 0.150\% \\
		0.075\% & 0.150\% & 0.175\% \\
		0.100\% & 0.100\% & 0.200\% \\
		0.125\% & 0.050\% & 0.225\% \\
		0.140\% & 0.020\% & 0.240\% \\
		0.145\% & 0.010\% & 0.245\% \\
	\end{tabular}
	\label{table1}
\end{table}

\section{Simulation and results}\label{Simulation and results}
In this study, we checked the market impact and volatility by changing the market maker rebate as shown Table \ref{table1}.We set the initial values of the model parameters as follows: $n=990$, $m=10$, ${w_1}_{max}=1$, ${w_2}_{max}=10$, $u_{max}=1$, $\tau_{max}=10{,}000$, $\sigma_{\epsilon}=0.06$, $Est=0.003$, $\Delta P=1.0$, $P_f=10{,}000$, 
$t_l=10{,}000$, $t_c=20{,}000$, $k_l=4.0$, $\delta _l=0.01$, and $w_{pm}=0.00000005$.
We ran simulations from $t = 0$ to $t = 1{,}000{,}000$ and defined the end time of simulations as $t_e$.

We now introduce the market impact $MI$, defined as how much higher
the price at which algorithm agents bought an asset is than the fundamental price $P_f$:
\begin{equation} 
	MI=\frac{1}{n_{{\it buy}}}\sum_{j=1}^{n_{{\it buy}}}\frac{P^{t^j_{{\it buy}}}-P_f}{P_f},
	\label{mi}
\end{equation}
where $n_{{\it buy}}$ is the amount of assets that algorithm agents bought and $P^{t^j_{{\it buy}}}$ is the price at which algorithm agents bought assets at time $t^j_{{\it buy}}$ 
The market impact is also used to calculate the average price at which algorithm agents bought the asset. 
The market price $P^t$ becomes almost the same as the fundamental price $P_f$ in the case without algorithm agents. Therefore, when orders of the algorithm agents do not have an impact on the price formation, $MI = 0$ \cite{MKKMIYY16}. A larger $MI$ indicates that the orders have more impact on price formation.

Volatility is defined as the standard deviation of the return $\ln(P^t/P^{t-1})$.

We define the market inefficiency $M_{\it ie}$ to measure the market efficiency as follows \cite{VDV13}:
\begin{equation} 
	M_{\it ie}=\frac{1}{t_e}\sum_{t=0}^{t_e}\frac{P^t-P_f}{P_f}.
	\label{mie}
\end{equation}
The market inefficiency is defined as the actual difference between market and fundamental prices. If the market was perfect efficiency, then the market prices would be exactly same as the fundamental price, i.e., $M_{\it ie} = 0$. A larger $M_{\it ie}$ indicates a more inefficient market. 

\subsection{Validation of artificial market model}\label{Validation of artificial market model}

\begin{table}[t]
	\begin{center}
		\caption{Stylized facts($\theta _{pm}=0.300\%$)\label{table2}}
		\begin{tabular}{cc | c} \hline
			Kurtosis &  & 17.54 \\ \hline
			 & Lag &  \\
			& 1 & 0.045 \\
			Autocorrelation & 2 & 0.045 \\
			coefficients & 3 & 0.044 \\
			for squared returns & 4 & 0.042 \\
			 & 5 &0.040 \\ \hline
		\end{tabular}
	\end{center}
\end{table}

As many empirical studies have mentioned \cite{Co01,Se06}, a fat tail and volatility clustering appear in actual markets, which are two stylized facts of financial markets. Therefore, we set the artificial market parameters so as to replicate these features. 
Table \ref{table2} shows the statistics for stylized facts in the case that the initial cash coefficient is 10 for which we calculated the price returns, i.e., $\ln({P^t/P^{t-1}})$, at intervals of 100 time units. As shown, both kurtosis and autocorrelation coefficients for squared returns with several lags are positive, which means that the runs for all five patterns replicate a fat tail and volatility clustering. This indicates that the model replicates long-term statistical characteristics observed in real financial markets. Since similar results were obtained for the other values of the initial cash coefficient, those results are omitted here due to space limitations.

A previous empirical study showed that kurtosis of returns and autocorrelation coefficients for squared returns with several lags are positive when a fat tail and volatility clustering appear in actual markets \cite{Ts05}. For example, the kurtoses of monthly log returns of U.S. bonds, the S\&P composite index of 500 stocks, and Microsoft are 4.86, 7.77, and 1.19, respectively. Note that kurtosis of returns depend on the kinds of financial assets. Likewise, autocorrelation coefficients for squared returns with several lags of the S\&P composite index of 500 stocks are 0.0536, 0.0537, 0.0537, 0.0538, and 0.0538 when lags are from 1 to 5, respectively. It seems that the results of our model do not differ significantly from these results. Therefore, it can be confirmed that the proposed model is valid.

\subsection{Results and discussion}\label{Results and discussion}
As a result of the simulations, we found that volatility, market impact, and market inefficiency decrease as the market maker rebate increases.

\subsubsection{Volatility}\label{volatility}
\begin{figure}[t]
	\includegraphics[width=\linewidth]{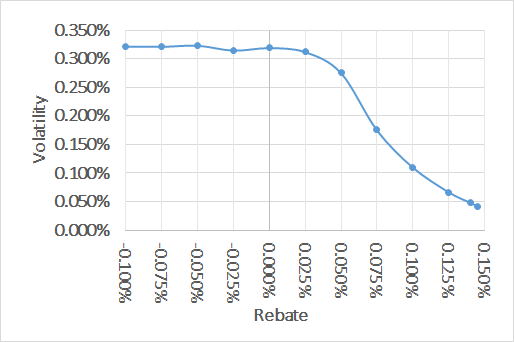}
	\caption{Volatility as a function of the market maker rebate\label{vola}}
\end{figure}

Fig. \ref{vola} shows volatility as a function of the market maker rebate. Volatility decreases when the market maker rebate increases. The reason for this result is as follows. As the market maker rebate increases, the maker can offer orders with a narrower spread, because the expected return of the market maker depends on both the market maker rebate and the spread (refer to Eq.~(\ref{basespread})). If the market maker narrows the spread between their orders, then it becomes easier for the normal agent's order to execute the market maker orders. Therefore, the market price tends to converge between the market maker's order prices and volatility decreases.

\subsubsection{Market impact}\label{Market impact}
\begin{figure}[t]
	\includegraphics[width=\linewidth]{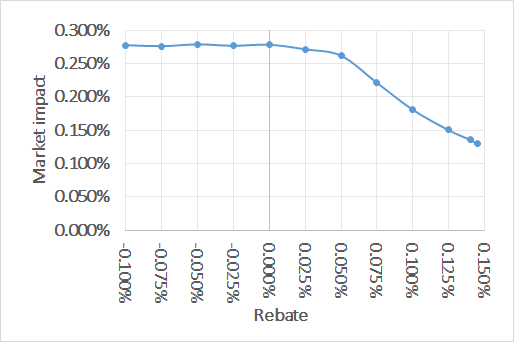}
	\caption{Market impact as a function of the market maker rebate\label{market_impact}}
\end{figure}

Fig. \ref{market_impact} shows the market impact as a function of the market maker rebate. The market impact decreases as the market maker rebate increases. For the reasons mentioned , the market maker can offer orders with a narrower spread. Therefore, as the buy market order prices of algorithm agents match the sell limit orders of the market maker, the market price does not rise much and the market impact decreases. 

\subsubsection{Market inefficiency}\label{Market inefficiency}
\begin{figure}[t]
	\includegraphics[width=\linewidth]{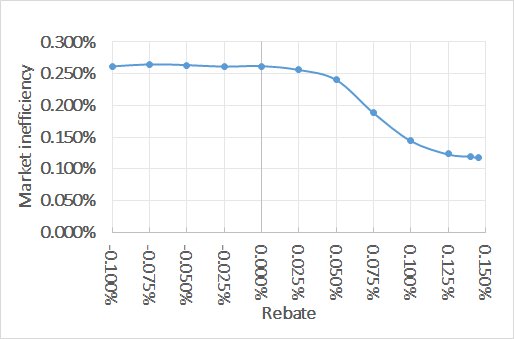}
	\caption{Market inefficiency as a function of the market maker rebate\label{market_inefficiency}}
\end{figure}

Fig. \ref{market_inefficiency} shows the market inefficiency as a function of the market maker rebate. In the market with the maker-taker fee structure, the market maker offers orders with a narrower spread around the fundamental price. As the taker's orders match the orders of the market maker, the market price converges to the fundamental price. Therefore, market inefficiency decreases, that is, market efficiency increases, when the market maker rebate increases. 

\subsubsection{Total cost of a taking order}\label{Total cost of a taking order}
\begin{figure}[t]
	\includegraphics[width=\linewidth]{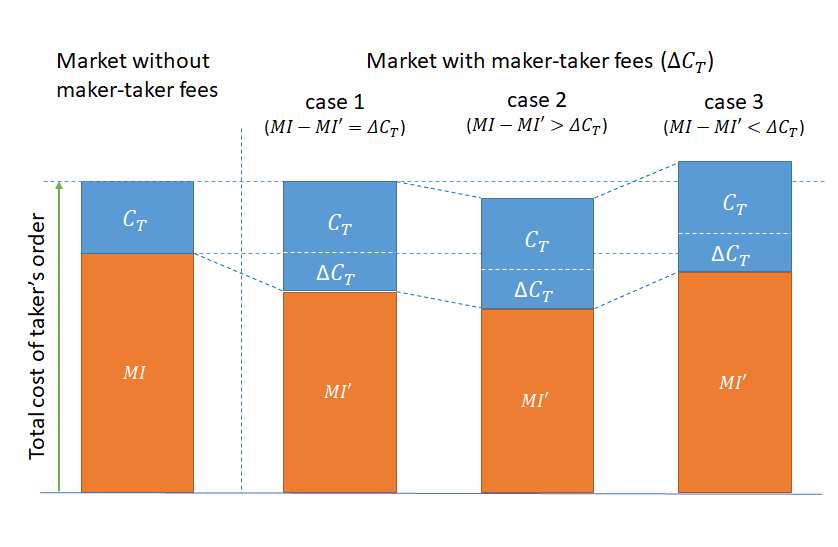}
	\caption{Relationship between the market impact and the increment of the taker fee\label{MIcases}}
\end{figure}

Fig. \ref{MIcases} shows some cases of the total cost of a taking order. 
Generally, if the decrement of the market impact due to the maker-taker fees is larger than the increment of the taker fee due to the maker-taker fees, then the total cost of a taking order in the market with the maker-taker fee structure is lower than that of a taking order in the market without it (refer to case 2 in Fig. \ref{MIcases}). 
This means that the algorithm agents in the market with the maker-taker fee structure can trade at lower transaction costs than those in the market without it. 
Note that $\Delta C_T$ is the increment of the taker fee due to the maker-taker fees and $MI'$ is market impact due to the maker-taker fees.
On the other hand, the algorithm agents in the market with the maker-taker fee structure have to trade at higher transaction costs than those in the market without it, if the decrement of the market impact due to the maker-taker fees is less than the amount of the increment of the taker fee (refer to case 3 in Fig. \ref{MIcases}).

\begin{figure}[t]
	\includegraphics[width=\linewidth]{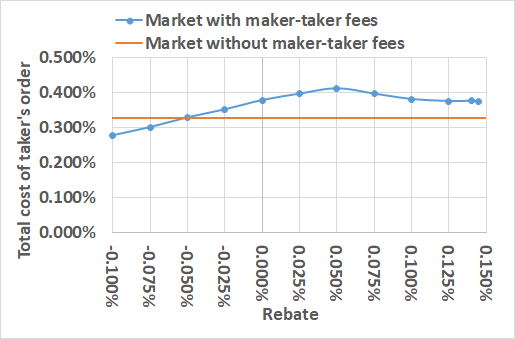}
	\caption{Total cost of the taking orders as a function of the market maker rebate\label{total_cost}}
\end{figure}

Fig. \ref{total_cost} shows a comparison of the total cost of a taking order between with and without the maker-taker fee structure in the market. 
Note that Fig. \ref{total_cost} shows the total cost of a taking order expressed as its ratio to the sum of it and the trading price.
The total cost of a taking order in the market without the maker-taker fee structure is 0.327\% (orange line in Fig. \ref{total_cost}). On the other hand, the total cost of a taking order in the market with the maker-taker fee structure is generally higher than that in the market without the maker-taker fee structure.

Therefore, we find that the maker-taker fee structure may create a disadvantageous environment for algorithm agents to trade in.

\section{Conclusion}\label{Conclusion}
In this study, we investigated the effect of maker-taker fees on the market from the viewpoints of volatility, market impact, and market efficiency by using an artificial market model. Furthermore, we also checked the relationship between the total costs of the taker's trade and market efficiency. As a result, we found that maker-taker fees contributed to decreases in volatility and market impact and an increase in market efficiency.  However, we also confirmed that it causes an increase in the total cost of the taking orders. 
Thus, if an exchange charges a high maker-taker fee, market makers are likely to make a profit, while other investors may leave the market due to higher transaction costs. 
Therefore, regulators should make regulations regarding the maker-taker fee structure in consideration of the above points, or at least avoid regulations that encourage exchanges to rebate market makers.

In our future work, we plan to confirm the effect of maker-taker fees on a market in which algorithm agents who always place not a market buy order but a market sell order for one share participate.

\section*{DISCLAIMER}
It should be noted that the opinions contained herein are solely those
of the authors and do not necessarily reflect those of SPARX Asset
Management Co., Ltd.
\begin{acks}
This work was supported by JSPS KAKENHI Grant Number 20K04977.
\end{acks}

\bibliographystyle{ACM-Reference-Format}
\bibliography{btxj17}

\end{document}